# Dark-field Tomography: Modeling and Reconstruction

W. Cong, F. Pfeiffer, M. Bech, O. Bunk, C. David, and G. Wang

**Abstract:**

Dark-field images are formed by small-angle scattering of x-ray photons. The small-angle scattering signal is particularly sensitive to structural variations and density fluctuation on a length scale of several ten to hundred nanometers, offering a new contrast mechanism to reveal subtle structural variation of object. In this paper, we derive a novel physical model to describe x-ray absorption and small-angle scattering, and use the proposed model to reconstruct the volumetric small-angle scattering images. The numerical experiments and test experiments demonstrate that the reconstructed scattering images reveal unique features with a high contrast resolution. The proposed approach has great potential in biomedical imaging, nondestructive detections, and other applications.

# 1. Introduction

The conventional x-ray computed tomography (CT) is based on x-ray attenuation, and yields sufficient contrast for substances with high density differences. However, this technique cannot achieve satisfactory sensitivity and specificity for low attenuation matters, such as soft biological tissues [1, 2]. In contrast to attenuation mechanism, small-angle scattering is particularly sensitive to structural variations and density fluctuation on a length scale of several ten to hundred nanometers. It provides a new contrast to reveal subtle structural variation of matters. Pfeiffer and coworkers proposed a grating interferometer technique to produce high quality dark-field images using a hospital-grade x-ray tube. The dark-field images of biological specimens present significantly higher contrast resolution than conventional attenuation-based images [3]. Recently, a single grating was used to modulate the x-ray beam. The modulated image contains a primary image and a grid harmonic image. The ratio between the harmonic and primary images reveals a pure scattering image [4]. Moreover, a fan-beam was proposed to illuminate an object slice for acquisition of coherent scattering data with multiline detectors [5]. The central detector row receives the transmitted radiation while the out-of-center rows record only scattered radiation. The technique is able to perform a rapid scanning for the object and provides a significant increment in image contrast for quantitative analyses.

Strobl *et al* proposed a method to simulate the broadening of the angular distribution of small angle scattering for dark field tomographic imaging. This broadening is related to both microscopic structure and multiple scattering along the path length through a matter [6, 7]. Harding et al directly applied the filtered backprojection algorithm to reconstruction the scattering contrast images from dark field data [5]. However, the propagation of x-ray photons through matter is a complex process, which experiences both absorption and scattering simultaneously. The scattering is caused by the changes in the refractive index, and the absorption depends on the density of matter. A photon propagation model describes photon interaction with matter, and is essential for tomographic imaging. In this paper, we derive a novel physical model to describe x-ray beam absorption and small-angle scattering. Then, we apply the proposed model to reconstruct the volumetric small-angle scattering images.

1. **X-ray small-angle scattering**

In quantum mechanics, light is considered with both wave and particle behaviors. Hence, the x-ray photon transmission can be treated as a beam of particles propagating through an object. As x-ray photons interact with the object, some photons would be deflected from the original direction due to a difference in the refractive index to generate a scattering signal. Thus, x-ray photons can be divided into transmitted photons traveling along a straight line in a direction $\theta$ and scattered photons deflected from the original direction. The propagation of transmitted photons along the direction $\theta$ can be well described by the Beer-Lambert law,

$$\Phi_a(\mathbf{r}_0 + R\theta) = \Phi_a(\mathbf{r}_0 + t\theta) \exp\left(-\int_t^R \mu_t(\mathbf{r}_0 + s\theta) ds\right) \qquad (1)$$

where $\Phi_a(\mathbf{r}_0 + t\theta)$ is the light intensity along the direction $\theta$, and $\mu_t$ the attenuation coefficient defined as a sum of absorption coefficient $\mu_a$ and scattering coefficient $\mu_s$, that is $\mu_t = \mu_a + \mu_s$. Eq. (1) can be reduced to a Radon transform,

$$\ln\left[\frac{\Phi_a(\mathbf{r}_0)}{\Phi_a(\mathbf{r}_0 + R\theta)}\right] = \int_0^R \mu_t(\mathbf{r}_0 + s\theta) ds \qquad (2)$$

where $\Phi_a(\mathbf{r}_0)$ and $\Phi_a(\mathbf{r}_0 + R\theta)$ express the intensity values from the x-ray source and upon the detector respectively after the x-ray propagation along the direction $\theta$ through the object. Based on the attenuation-based CT technology, the unknown x-ray attenuation coefficient $\mu_t$ can be reconstructed using the conventional reconstruction algorithms, such as the filtered backprojection (FBP) algorithm [8, 9].

While an x-ray beam propagates in matter along a straight line, some photons would experience a small-angle forward scattering. The so-called dark-field image is formed through the small-angle scattering of x-rays. The scattered photon intensity $\Phi_s(\mathbf{r})$ depends on both absorption and scattering coefficients of object. According to the energy conservation principle, the difference $d\Phi_s(\mathbf{r})$ of the scattered photon intensity between the opposite sides of an elementary volume with a cross sectional area $dA$ and length $dh$ along the direction $\theta$ is equal to the difference between intensity of scattered photon from primary beam subtracting the intensity of photons absorbed by matter, which can be expressed as follows,

$$d\Phi_s(\mathbf{r})dA = \Phi_a(\mathbf{r})dA\,\mu_s dh - \Phi_s(\mathbf{r})dA\,\mu_a dh. \tag{3}$$

Since $d\Phi_s(\mathbf{r})dA = \theta \cdot \nabla \Phi_s(\mathbf{r})dhdA$, a differential equation in terms of the scattered photon intensity can be obtained from Eq. (3):

$$\theta \cdot \nabla \Phi_s(\mathbf{r}) + \mu_a \Phi_s(\mathbf{r}) = \mu_s \Phi_a(\mathbf{r}), \tag{4}$$

where $\mu_a \Phi_s(\mathbf{r})$ represents the loss of the scattering intensity due to absorption, and $\mu_s \Phi_a(\mathbf{r})$ is the quantity of scattered photons from the primary beam $\Phi_a(\mathbf{r})$. In other words, Eq. (4) describes the balance of the photons between the input and output of an elementary volume at the given direction $\theta$. Because Eq. (4) is a linear first-order differential equation, its solution can be obtained in the closed form:

$$\Phi_s(\mathbf{r}_0 + R\theta) = \int_0^R \mu_s(\mathbf{r}_0 + t\theta)\Phi_a(\mathbf{r}_0 + t\theta)\exp\left(-\int_t^R \mu_a(\mathbf{r}_0 + s\theta)ds\right)dt. \tag{5}$$

Substituting Eq. (1) into Eq. (5), we obtain

$$\Phi_s(\mathbf{r}_0 + R\theta) = \Phi_a(\mathbf{r}_0 + R\theta) \int_0^R \mu_s(\mathbf{r}_0 + t\theta) \exp\left(\int_t^R \mu_s(\mathbf{r}_0 + s\theta) ds\right) dt. \qquad (6)$$

Using a variable transformation, Eq. (6) can be simplified to a Radon transform with respect to the scattering coefficient distribution:

$$\ln\left[1 + \frac{\Phi_s(\mathbf{r}_0 + R\theta)}{\Phi_a(\mathbf{r}_0 + R\theta)}\right] = \int_0^R \mu_s(\mathbf{r}_0 + s\theta) ds \qquad (7)$$

where $\Phi_a(\mathbf{r}_0 + R\theta)$ and $\Phi_s(\mathbf{r}_0 + R\theta)$ are the intensity of transmission photon and the intensity of small-angle scattering photons on the detectors, respectively. Eq. (7) describes the relationship between the scattering characteristic of matter and measured photon transmission and small-angle scattering data. Eq. (7) is also a standard Radon transform for the scattering coefficient, so classical reconstruction algorithms, such as filtered backprojection (FBP) algorithm, algebraic reconstruction technique (ART), can be applied to reconstruct the scattering coefficient distribution tomographically [8, 9].

## 2. Numerical simulations

### 2.1. Shepp-Logan's phantom

We first employed Shepp-Logan's phantom to evaluate the proposed dark-field tomographic imaging method [9]. The phantom parameters were listed in Table I. Ten ellipses was included in the phantom to mimic the subtle structures of the head tissues. In Table I, $a$ and $b$ are the semi-axes of an ellipse, $x_0$ and $y_0$ specify the center of an ellipse, $\mu_t$ denotes an x-ray attenuation coefficient, and $\mu_s$ the x-ray scattering coefficient, and $\phi$ the angle (in degrees) between the horizontal semi-axis of the ellipse and the x-axis of the reconstruction system. The phantom had a less density variations and low attenuation contrast. The projection data of the dark-field and bright-field at different positions and angles were computed using our in-house Monte Carlo simulator. The number of projections was 180. The reconstruction matrix was set to 512 by 512. Then, the FBP algorithm based on the proposed scattering model was implemented to reconstruct the scattering and attenuation coefficient distributions, respectively. As a result, the reconstructed scattering coefficient images showed a high contrast resolution for different tissue structures, as shown in Fig.1 (a) with interfaces highlighted due to dark-field imaging. In contrast, the conventional attenuated-based reconstruction shown in Fig.1 (b)

*Table I. Parameters of the Shepp-Logan Phantom*

| No. | $\mu_t$ | $\mu_s$ | $a$ | $b$ | $x_0$ | $y_0$ | $\phi$ |
|---|---|---|---|---|---|---|---|
| 1 | 0.06 | 0.0020 | 0.6900 | 0.9200 | 0 | 0 | 0 |
| 2 | 0.04 | 0.0020 | 0.6624 | 0.8740 | 0 | -0.0184 | 0 |
| 3 | 0.02 | 0.0025 | 0.1100 | 0.3100 | 0.22 | 0 | -18 |
| 4 | 0.02 | 0.0025 | 0.1600 | 0.4100 | -0.22 | 0 | 18 |
| 5 | 0.01 | 0.0015 | 0.2100 | 0.2500 | 0 | 0.35 | 0 |
| 6 | 0.05 | 0.0035 | 0.0460 | 0.0460 | 0 | 0.1 | 0 |
| 7 | 0.05 | 0.0035 | 0.0460 | 0.0460 | 0 | -0.1 | 0 |
| 8 | 0.01 | 0.0025 | 0.0460 | 0.0230 | -0.08 | -0.605 | 0 |
| 9 | 0.01 | 0.0025 | 0.0230 | 0.0230 | 0 | -0.606 | 0 |
| 10 | 0.01 | 0.0025 | 0.0230 | 0.0460 | 0.06 | -0.605 | 0 |

could not discern the different types of tissues because of an insufficient attenuation contrast.

**3.2. Breast phantoms**

The proposed dark-field tomographic imaging technique was also numerically evaluated using the digital breast phantom [10]. The breast phantom represents an uncompressed breast of a half ellipsoidal shape, containing anatomical and pathological features of different sizes and contrasts. The three ellipsoidal semiaxes were set to 50, 50, and 100 mm. The skin thickness was set to 2.5 mm. The fibroses were modeled as cylinders, the calcifications and mass as balls, as shown in Fig.2. The phantom was positioned in the nonnegative space, attached to the chest wall defined on z =0. Masses, Fibroses and calcifications were centered on the planes of z =30 mm and z=50 mm, respectively. Table II-IV lists the geometrical parameters, attenuation and scattering properties at 38 keV of the masses, fibroses, and calcifications, respectively [10]. The breast phantom was of low attenuation contrast between different breast tissues. We adopted the parallel-beam mode to scan the phantom, and simulated the dark-field and bright-field data at different positions and angles using our x-ray Monte Carlo simulator. Then, the reconstruction algorithm was again used to reconstruct the scattering and attenuation images. Similarly, the reconstructed scattering images had an excellent contrast to differentiate the tissue structures. Fig.3 (a) and Fig.4 (a) showed the reconstructed images at the cross sections of z=30mm and z=50mm, respectively. In contrast, the conventional attenuated-based reconstructions shown in Fig.3 (b) and Fig.4 (b) could not discriminate the tumors from the normal breast tissues.

*Table II. Dimensions, attenuation and scattering properties of masses.*

| Mass No. | 1 | 2 | 3 | 4 | 5 |
|---|---|---|---|---|---|
| Radius (mm) | 3 | 2 | 1 | 3 | 1 |
| $\mu_t$ | 0.023 | 0.025 | 0.030 | 0.031 | 0.028 |
| $\mu_s$ | 0.0023 | 0.0028 | 0.0035 | 0.0038 | 0.0030 |

*Table III. Dimensions, attenuation and scattering properties of fibroses.*

| Fibrous No. | 1 | 2 | 3 | 4 | 5 |
|---|---|---|---|---|---|
| Radius (mm) | 1 | 1 | 1 | 1 | 1 |
| Height (mm) | 10 | 10 | 10 | 10 | 10 |
| $\mu_t$ | 0.012 | 0.015 | 0.015 | 0.020 | 0.020 |
| $\mu_s$ | 0.0018 | 0.0020 | 0.0020 | 0.0025 | 0.0025 |

*Table IV. Dimensions, attenuation and scattering properties of fibroses.*

| Calcification No. | 1 | 2 | 3 | 4 | 5 |
|---|---|---|---|---|---|
| Radius (mm) | 0.5 | 0.7 | 0.7 | 0.5 | 0.5 |
| $\mu_t$ | 0.030 | 0.032 | 0.032 | 0.035 | 0.035 |
| $\mu_s$ | 0.0036 | 0.0042 | 0.0042 | 0.0045 | 0.0048 |

## 3. Test experiments

To test our theoretical model on experimental dark-field data obtained with a grating interferometer, we carried out some test experiments at the beamline ID19 of the European Synchrotron Radiation Facility (ESRF, Grenoble). A small beatle (ex-vivo) was used as a on a strongly scattering test sample. A monochromatic x-ray beam of 24.9 keV ($\lambda = 0.0498$ nm) was used for the measurements. The interferometer was placed at a distance of 140 m from the wiggler source (see [11] for more details on the grating parameters).

The field-of-view was matched to the size of the specimen and was $16.1 \times 16.1$ mm2. To achieve a very high angular, and thus phase sensitivity, the distance between G1 and G2 was chosen to be as large as 361 mm (ninth fractional Talbot distance). The images were recorded using a 15 μm thick polycrystalline gadolinium oxysulfide scintillation screen with a magnifying optical lens system and a cooled charge coupled device (CCD). We used the FReLoN 2000 (a fast-readout, low-noise CCD developed at the ESRF) with 1024 × 1024 pixels and a 28.0 × 28.0 μm$^2$ pixel size (in the 2 × 2 binning mode). Due to the magnifying lens system, the effective pixel size in the recorded images was 15.7 × 15.7 μm$^2$.

In total 721 projection angles over 180 degree were recorded. The Fringe-scanning method was used to acquire eight images $I_k(x,y)$ ($k=1,2,\cdots,8$) at every projection angle by shifting moiré fringes. In the Talbot interferometer, the shift was attained by displacing one of the gratings in the direction parallel to its diffraction vector. The transmission intensity and dark-field signals can be extracted from the measured intensity images respectively as follows,

$$a(x,y) = \sum_{k=1}^{M} I_k(x,y)$$

and

$$b(x,y) = \sqrt{\left(\sum_{k=1}^{M} I_k(x,y)\sin(kx_g)\right)^2 + \left(\sum_{k=1}^{M} I_k(x,y)\cos(kx_g)\right)^2}$$

where $kx_g$ is the displacement of the grating. The number of projection angles is 721. The dark-field tomography method was then applied to reconstruct the beetle tissue structure from the acquired dark field signals. Fig. 5(a) and Fig. 6(a) show dark-field tomographic images. It can be observed that the new contrast mechanism helped identify structural features of the investigated sample. The small-angle-scattering–based reconstructions as derived from the dark-field data are particularly useful for identifying structures in an object on the scale of about a hundred nanometers to a few micrometers.

For comparison, the corresponding attenuation-based tomographic slices are shown in Fig.5 (b) and Fig.6 (b), which exhibit a low attenuation contrast of the biological sample.

## 4. Discussion and conclusions

In summary, we have developed a novel physical model to describe both x-ray attenuation and small-angle scattering. The numerical and biological experiments have shown that dark-field tomographic imaging can reveal detailed structural variation of matter, producing a higher contrast resolution for low attenuation contrast features than conventional attenuation-based computerized tomography. Dark-field images are particularly sensitive and specific to boundaries and interfaces in the matter, producing strong dark-field signal to reveal detailed structural information of matter. Additionally, the radiation dose strongly relies on x-ray absorption properties of matter. The probability of x-ray photoelectric absorption drops off rapidly as a function of the incident X-ray photon energy. Higher x-ray photon energy produces lower radiation absorption, resulting in a poor contrast for low absorption matters. Hence, attenuation-based CT often uses lower x-ray photon energy to enhance contrast resolution for low attenuation media, inducing a considerable radiation dose. In contrast, the principal advantage of our dark-field imaging method is that the contrast of small-angle scattering imaging does not depend solely on photon absorption, so x-ray energies can be chosen to minimize radiation absorption in matters. The proposed approach has a great potential for a wide range of applications, including clinical and pre-clinical imaging, food inspection, security screening, and industrial non-destructive testing.

**Acknowledgement:** G.W. and W. Cong acknowledge the support from the National Institutes of Health (Grants CA 135151, EB006036, EB008476, and CA127189) and Toshiba Medical Systems (Advanced Imaging Methods Grant). We gratefully acknowledge the assistance of C. Kottler and P. Cloetens in the experiments.

**References:**

1. Momose, A., et al., *Phase-contrast X-ray computed tomography for observing biological soft tissues (vol 2, pg 473, 1996).* Nature Medicine, 1996. **2**(5): p. 596-596.

2. Chapman, D., et al., *Diffraction enhanced x-ray imaging.* Physics in Medicine and Biology, 1997. **42**(11): p. 2015-2025.

3. Pfeiffer, F., et al., *Hard-X-ray dark-field imaging using a grating interferometer.* Nature Materials, 2008. **7**(2): p. 134-137.

4. Wen, H., et al., *Spatial harmonic imaging of X-ray scattering initial results.* IEEE Transactions on Medical Imaging, 2008. **27**(8): p. 997-1002.

5. Harding, G., *X-ray scatter tomography for explosives detection.* Radiation Physics and Chemistry, 2004. **71**(3-4): p. 869-881.

6. Strobl, M., W. Treimer, and A. Hilger, *Small angle scattering signals for (neutron) computerized tomography.* Applied Physics Letters, 2004. **85**(3): p. 488-490.

7. Strobl, M., et al., *Neutron dark-field tomography.* Physical Review Letters, 2008. **101**(12): p. -.

8. Natterer, F., *The mathematics of computerized tomography.* 1986, Stuttgart Chichester: B.G. Teubner ; Wiley. x, 222.

9. Kak, A.C. and M. Slane, *Principles of Computerized Tomographic Imaging.* 1988, New York: IEEE Press.

10. Zeng, K., et al., *Cone-beam mammo-computed tomography from data along two tilting arcs.* Medical Physics, 2006. **33**(10): p. 3621-3633.

11. Pfeiffer, F. et al., *High-resolution brain tumor visualization using three-dimensional x-ray phase contrast tomography,* Physics in Medicine and Biology, 2007. **52**(23), p 6923-6930.

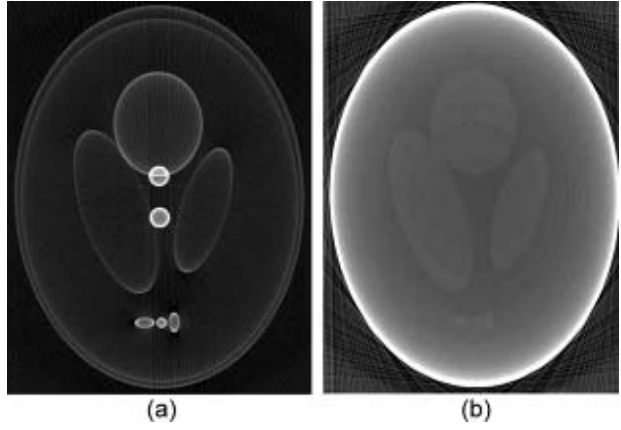

Fig.1. Comparison between dark-field tomography and conventional CT. Image reconstructed from (a) the dark-field data and (b) the transmission data.

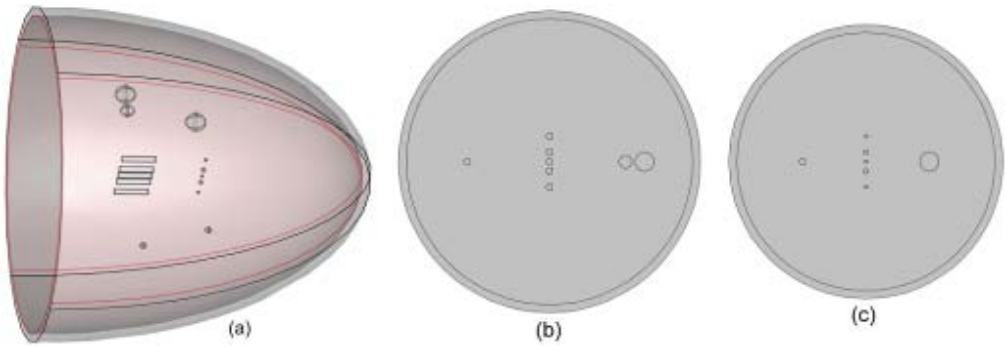

Fig.2. 3D breast phantom. (a) A volumetric rendering, (b) a cross section at z=30mm, and (c) a cross section at z=50mm.

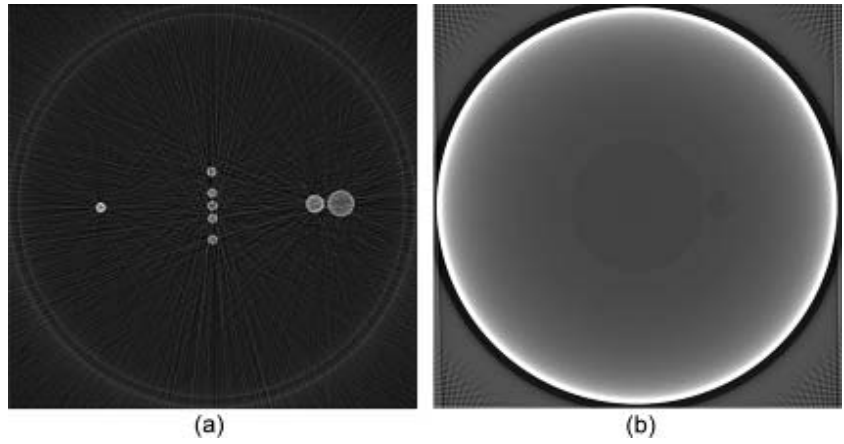

Fig.3. Comparison between dark-field tomography and conventional CT for the breast phantom, image at z=30mm slice reconstructed from (a) the dark-field data and (b) the transmission data.

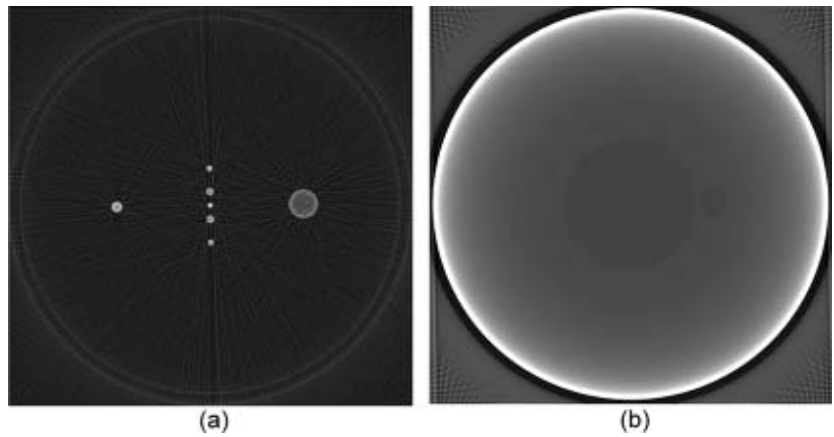

Fig. 4. Comparison between dark-field tomography and conventional CT for the breast phantom, images at z=50mm slice reconstructed from (a) dark-field data and (b) transmission data.

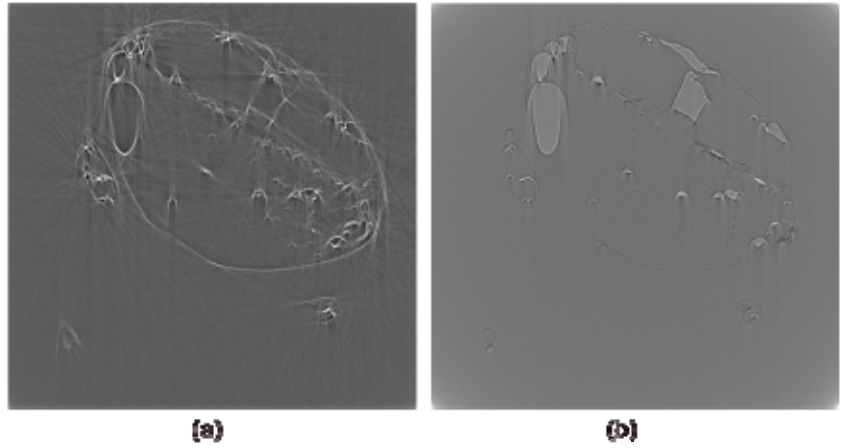

Fig. 5. Comparison between dark-field tomography and conventional CT for Beatle. Images at 270th slice reconstructed from (a) dark-field data and (b) transmission data.

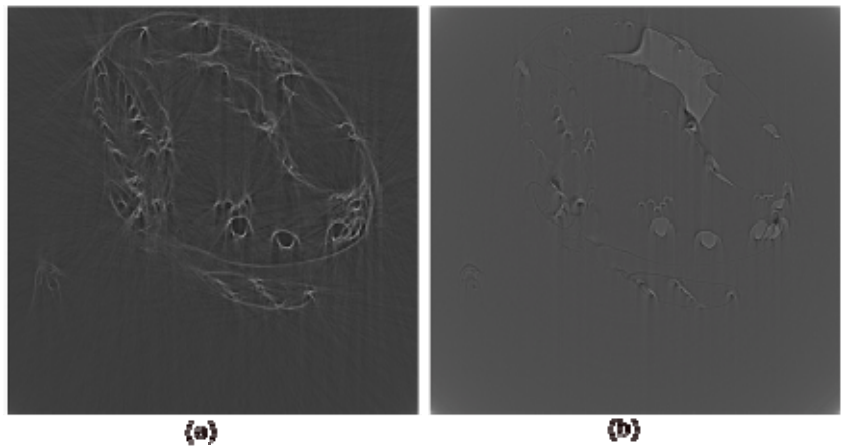

Fig. 6. Comparison between dark-field tomography and conventional CT for Beatle. Images at 330th slice reconstructed from (a) dark-field data and (b) transmission data.